\documentclass[twocolumn,preprintnumbers,amsmath,amssymb]{revtex4}  

\usepackage{graphicx}
\usepackage{dcolumn}
\usepackage{bm}
\usepackage{multirow}
\usepackage{color}
\usepackage{amsmath}

\begin{document}

\title{Control of Exciton Valley Coherence in Transition Metal Dichalcogenide Monolayers}

\author{G. Wang$^1$}
\author{X. Marie$^1$}
\author{B. L. Liu$^2$}
\email{blliu@iphy.ac.cn}
\author{T. Amand$^1$}
\author{C. Robert$^1$}
\author{F. Cadiz$^1$}
\author{P. Renucci$^1$}
\author{B. Urbaszek$^1$}
\email{urbaszek@insa-toulouse.fr}
\affiliation{%
$^1$ Universit\'e de Toulouse, INSA-CNRS-UPS, LPCNO, 135 Av. Rangueil, 31077 Toulouse, France}
\affiliation{%
$^2$ Beijing National Laboratory for Condensed Matter Physics, Institute of Physics, Chinese Academy of Sciences, P.O. Box 603, Beijing 100190, People’s Republic of China}

\begin{abstract}
The direct gap interband transitions in transition metal dichalcogenides monolayers are governed by chiral optical selection rules. Determined by laser helicity, optical transitions in either the $K^+$ or $K^-$ valley in momentum space are induced. Linearly polarized laser excitation prepares a coherent superposition of valley states.  Here we demonstrate the control of the exciton valley coherence in monolayer WSe$_2$ by tuning the applied magnetic field perpendicular to the monolayer plane. We show rotation of this coherent superposition of valley states by angles as large as 30 degrees in applied fields up to 9~T.  This exciton valley coherence control on ps time scale could be an important step towards complete control of qubits based on the valley degree of freedom.  

\end{abstract}

                             
\maketitle
Atomically thin layers of Van der Waals bonded materials open up new possibilities for fundamental physics in 2D systems and for new applications \cite{Geim:2013a,Novoselov:2005a,Castellanos:2016a}. Here the group-VI Transition metal dichalcogenides (TMDCs) of the form MX$_2$, where M=$Mo,W$ and X=$S,Se$ stand out: These indirect semiconductors in bulk form become direct semiconductors when thinned down to one monolayer (ML) \cite{Mak:2010a,Splendiani:2010a, Eda:2011a,Zhao:2013b,amani:2015a,Sundaram:2013a, Pospischil:2014a,Withers:2016,Conley:2013a}. 
The valence and conduction band extrema of a ML reside at the $K$ point of the Brillouin zone. Current research interest is stimulated by their strong light-matter interaction and the possibility to use the valley index as an information carrier and for exciting fundamental physics experiments \cite{Xiao:2012a,Xu:2014a,Mak:2014a,yang:2015a,Hao:2015a,Mak:2016a}. 
Due to the reduction of dielectric screening and the large effective carrier masses in TMDCs monolayers the light-matter interaction is dominated by the excitons (Coulomb bound electron-hole pairs), with binding energies up to several hundred meV \cite{Chernikov:2014a,Zhu:2015b,Ugeda:2014a,Wang:2015b,He:2014a,Hanbicki:2015a,Ye:2014a}. Using $\sigma^+ (\sigma^-)$ polarized excitation, the optical excitation of carriers in the $K^+$ ($K^-$) valley results in formation of an exciton with pseudo-spin $|+1\rangle$ ($|-1\rangle$) \cite{Mak:2012a,Zeng:2012a,Cao:2012a}. \\
\indent A basic requirement for quantum information processing experiments using the valley degree of freedom is the ability to completely control the state of a single qubit, as demonstrated for electron spins in quantum dots \cite{Press:2008a}. 
A universal single qubit gate is realized by a rotation of a single spin, for example, by any angle about an arbitrary axis. 
One strategy for qubit manipulation is to use picosecond or femtosecond laser pulses, permitting an arbitrary rotation to be completed within one spin precession period. 
Here a first important step was the demonstration in ML WSe$_2$ of the \textit{superposition} of two valley states. 
This is achieved optically by linearly polarized excitation, which results in strongly linearly polarized neutral exciton (X$^0$) emission \cite{Jones:2013a,Wang:2014b,Wang:2015g}. Contrary to coherent exciton manipulation in the model system of GaAs quantum well excitons \cite{Barad:1991a,Amand:1997a,Marie:1997a}, here in ML TMDCs the optical excitation can be at much higher energy than the transition \cite{Wang:2015g} i.e. strictly resonant excitation/detection of the coherent exciton states is not required. Despite these favorable conditions, so far a demonstration of exciton valley coherence control is lacking. \\
\begin{figure*}[tb]
\includegraphics[width=0.87\textwidth]{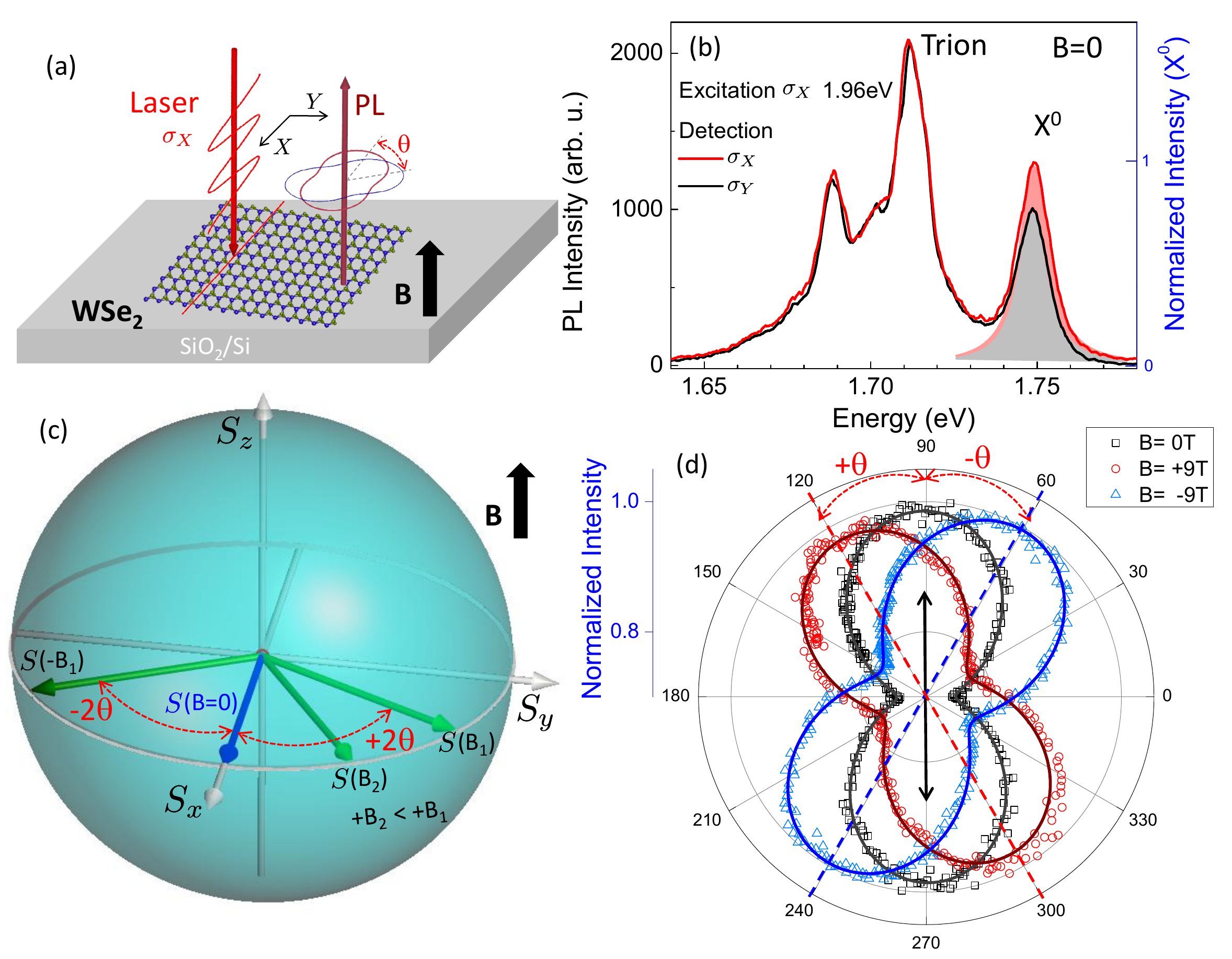}
\caption{\label{fig:fig1} \textbf{(a)} Schematic of experimental configuration. \textbf{(b)} Valley coherence generation following linearly polarized excitation $\sigma_X$ at 1.96~eV probed by through PL emission of X$^0$ at the typical energy 1.75~eV \cite{Wang:2014b} at T=4~K. The trion emission and the low energy emission probably linked to localized states are as expected unpolarized \cite{Jones:2013a} \textbf{(c)} Representation of valley coherence (average exciton pseudo-spin) on the Bloch sphere. At B=0, no rotation around the equator occurs as shown by the blue arrows. For $B\neq0$ the exciton pseudo spin precesses around the equator during the exciton lifetime. The green arrows correspond to the average pseudo spin states for $B\neq0$ governed by eq.~(1) and (2). By changing the amplitude and the direction of the magnetic field in the $Z$ direction the valley coherence can be tuned to different points on the equator, compare $S(B_1)$ with $S(B_2)$. \textbf{(d)} The normalized X$^0$ angle dependent intensity polar plots for B=0 (black), B=+9~T (red) and B=-9~T (blue). The normalized intensity 0.7 corresponds to the center and 1 to the outermost gray circle. The laser polarization direction (constant for all B-field values) is indicated by the black arrow.
}
\end{figure*}
\indent In this work we demonstrate that the neutral exciton valley coherence in monolayer WSe$_2$ can be controlled by an external magnetic field applied vertically to the sample plane. In the absence of external fields, the electronic states related by time reversal in the $K^+$ and $K^-$ valleys are degenerate. The valley exciton degeneracy can be lifted by a longitudinal magnetic field \cite{Li:2014a,srivastava:2015,Macneill:2015a,Aivazian:2015a,Arora:2016a,Wang:2015d} or the optical Stark effect \cite{kim:2014a,sie:2015b}. 
In our experiment the external magnetic field lifts the valley degeneracy and results in a change of the oscillation frequency of the coherent superposition of valley states. This corresponds to a rotation of valley coherence (i.e. the exciton pseudo-spin) and we clearly measure this rotation in our experiments with angles up to 30 degrees at $B=9$~T. This type of quantum beat process was observed initially in atoms and molecular systems \cite{Shoemaker:1974} and then intensely investigated for excitons in low dimensional GaAs structures by time-resolved optical techniques \cite{Barad:1991a,Amand:1997a,Marie:1997a}. Whereas in other material systems coherent manipulation is necessarily a 2 pulse experiment to read and write the quantum state, we show here that in ML TMDCs these experiments can be carried out with simple cw excitation and photo-luminescence (PL) detection. \\
\begin{figure}[ht!]
\includegraphics[width=0.45\textwidth]{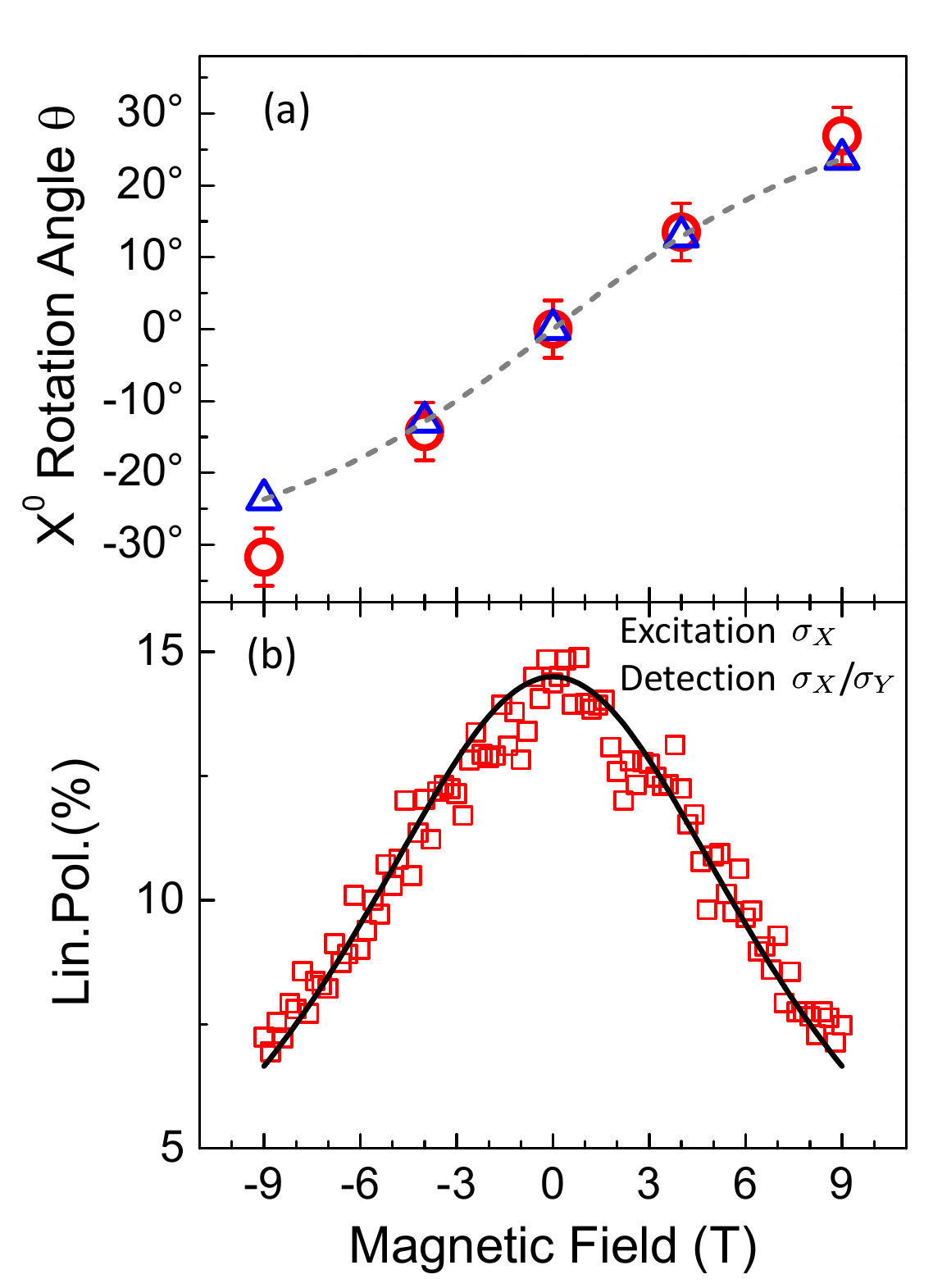}
\caption{\label{fig:fig2}  \textbf{(a)} Absolute rotation angle of linear polarization of X$^0$ PL for different magnetic field values (red open circles). The gray dashed curve is the calculated using $\theta=\frac{\mathrm{arctan}(\Omega/T_{S2}^*)}{2}$ and they are stressed by the blue open triangles at the experimental magnetic field for comparison. \textbf{(b)} The linear polarization of X$_0$ measured along fixed direction X and Y. The black solid line is the calculated value. Both gray dash line, the blue opened triangle in \textbf{(a)} and the black solid line in \textbf{(b)} are calculated by $g$=-3.7, $T^*_2$=0.37ps.} 
\end{figure} 
\textit{Samples and Experimental Set-up.---}
The WSe$_2$ ML flakes are prepared by micro-mechanical cleavage of a bulk crystal (from 2D Semiconductors) on SiO$_2$/Si substrates. The experiments are carried at T=4 K and in magnetic fields up to $\pm$ 9 T in Faraday configuration in a confocal microscope as shown in Fig.~1a. The excitation/detection spot diameter is $\approx1\mu$m, i.e. smaller than the typical ML diameter. The WSe$_2$ ML is excited by a linearly polarized ($\sigma_X$) continuous wave He-Ne laser (1.96~eV) to generate valley coherence (i.e. optical alignment of excitons \cite{Meier:1984a}). Our target is to detect the neutral exciton X$^0$ valley coherence in the linear basis in PL emission, see Fig.~1b. A liquid crystal based linear polarization rotator is applied in the detection path, to detect a possible rotation of the linear basis of the PL signal with respect to the initial linear excitation basis. This approach avoids any macroscopic mechanical movement during the measurement and gives an accurate map of the angle dependent PL intensity as schematically illustrated in Fig.~1a, the full data set is plotted in Fig.~1d. The PL signal is dispersed in a spectrometer and detected with a Si-CCD camera. Based on these time integrated PL results we can then generate the polar plot of the intensity of the X$^0$ emission for different magnetic field values and then monitor the rotation of the valley coherence by an angle $\theta$ as illustrated in Fig.~1a. Faraday effects of the optical set-up in applied fields have been systematically calibrated for plotting the valley coherence rotation angle. \\
\textit{Results and Discussion.---}
Our experiment consists of 3 steps: First we want to optically initialize a coherent superposition of exciton states. Second, during the exciton lifetime we want to rotate the exciton pseudo-spin. Third, we read out the final state after rotation. This 3 step experiment will of course only be successful if the coherence time is sufficiently long compared to the read-out time  \cite{Hao:2015a}.
In the experiment we monitor the neutral exciton X$^0$ PL emission, which is linearly polarized along the same axis as the excitation laser. This corresponds to successful valley coherence generation \cite{Jones:2013a}, as shown in Fig.~1b. 
The linear polarization of X$^0$ measured along the initial excitation direction $X$ (i.e. the laser polarization axis) is around 15\%. In contrast, both trion and lower energy emission peaks present no linear polarization as reported commonly for ML WSe$_2$ \cite{Jones:2013a,Wang:2014b}. In analogy to spin, the average of this coherent superposition of valley states can be represented by a vector that lies on the equator of a Bloch sphere as illustrated in Fig.~1c. 
The north $+S_z$ (south $-S_z$) pole on the Bloch sphere corresponds to a $\left|+1 \right\rangle$ ($\left|-1\right\rangle$) exciton states, which can be optically generated by $\sigma^+$ ($\sigma^-$) optical excitation. The equator corresponds to an X$^0$ in-plane valley pseudo-spin, coherent superposition of $\left|+1 \right\rangle$ and $\left|-1\right\rangle$ with different phases. 
Radiative recombination of the  X$^0$ results in photon emission. Importantly, the exciton's pseudo-spin direction in the equator will determine the linear polarization of the photon. In the absence of external fields, the linear polarization basis of the emitted photon is the same as the excitation laser i.e. no pseudo-spin rotation occurs during the short PL emission time $\tau\approx1$~ps \cite{Palummo:2015,Robert:2016a,pollmann:2015a,Korn:2011a} as the main axis of the polar plot Fig.~1d (black squares) is aligned along the laser polarization direction. \\
\indent In order to rotate the exciton spin, we now apply a magnetic field $B$ which lifts valley degeneracy.  At $B\neq0$ the exciton valley coherence state after a linearly polarized excitation along the $X$ direction evolves with time as $\left|X\right\rangle=\frac{1}{\sqrt{2}}(\left|+1\right\rangle e^{-i \Omega t/2}+\left|-1 \right\rangle e^{i \Omega t/2})$, where $\left|\pm1 \right\rangle$ are the exciton spin states, the energy difference between the two $K$ valleys $\hbar \Omega=g \mu_B B $ with $\mu_B$ the Bohr magneton and $g$ the X$^0$ Lande $g$-factor. The PL emission time in ML WSe$_2$ is of the order of $\tau \approx 1$~ps. It should therefore be possible to read out the final state after rotation within the first precession period by simply analysing the linear polarization basis of the X$^0$ PL. Clear rotations of the linear polarization axis are observed in Fig.~1d for B=-9~T and also B=+9~T. At B=+9~T we measure a rotation of $\theta=27\pm4$ degrees with respect to the initial laser excitation polarization. Changing from +9 to -9~T, we observe that $\theta$ also changes sign while keeping the same amplitude, as the exciton pseudo-spin rotation now occurs in the opposite direction. These trends are confirmed for the intermediate values at B=$\pm4$~T as plotted in Fig.~2a. Very similar data has been obtained for other ML WSe$_2$ samples.
The solid lines in Fig.1d are fits using $r=A_0+A_1 \mathrm{cos}[2(x-\theta)]$, where $x$ is the detection angle and $\theta$ represents the X$^0$ linear polarization angle. Experimentally, it is clear from Figs.~1d and 2a that we can control the rotation of the linearly polarized exciton emission with respect to the initial laser excitation axis.\\
\indent Now we give a very simple interpretation of the experiments in terms of rotating a superposition of exciton $\left|+1 \right\rangle$ and $\left|-1\right\rangle$ states originating from exciton pseudo-spins: In principle, in a finite magnetic field $B$ applied perpendicular to the ML, the valley coherence has a nonzero precession frequency $\Omega$ due to the energy difference $g \mu_B B $ between the $\left|+1 \right\rangle$ and $\left|-1\right\rangle$ exciton states. A certain time after the initialization of the valley coherence state, it will evolve to a new position in the equator of the Bloch sphere as shown in Fig.~1c by the green arrows labelled with $\pm B_1$. With a different magnetic field B$_2$ the final position of the valley coherence state can be selectively adjusted to a different rotation angle. Through the fits used in Fig.~1d, we deduce the magnetic field dependence of the rotation angle, shown in Fig.~2a with open red circles. In our simple approach we assume that the valley coherence generation rate does not change as a function of the applied magnetic field. This scenario is very similar to the standard Hanle effect \cite{Dyakonov:2008a}: the initial pseudo-spin generated along the $X$ direction (laser polarization axis) precesses around the magnetic field $B$ applied in the $Z$ direction (perpendicular to the ML plane) at frequency $\Omega=g \mu_B B/\hbar $. For linear excitation the generated pseudo-spin component $S_z=0$. In stationary conditions the pseudo-spin state in-plane components in an applied magnetic field B can be expressed as:
\begin{equation}
S_x(B)=\frac{S_x(0)}{1+(\Omega T_{S2}^*)^2}
\end{equation}
\begin{equation}
S_y(B)=\frac{ \Omega T_{S2}^*}{1+(\Omega T_{S2}^*)^2}S_x(0)
\end{equation}
where, $1/T_{S2}^*=1/\tau+1/T_{S2}$ , $\tau$ is the exciton lifetime, $T_{S2}$ is the exciton valley coherence time. The pseudo-spin rotates by an angle $\phi$, where tan$\phi=\Omega T_{S2}^*$. Therefore we can deduce the X$^0$ PL rotation angle $\theta=\phi/2$, as plotted in Fig.~2a (gray dashed line) using  $g=-3.7$ and $T_{S2}^*=0.37ps$. The values calculated at the  magnetic field values used in our experiment are plotted as blue open triangles for comparison with the data. This simple model yields excellent  agreement with our experimental results. The $g-$factor of X$^0$ in ML WSe$_2$ is extracted from circularly resolved magneto PL results \cite{Wang:2015d}, which is also similar to other reports for this material \cite{srivastava:2015, mitioglu:2015}. We therefore extract a value of $T_{S2}^*$ of around $0.37$~ps. Assuming an exciton lifetime of $\tau\approx2$~ps \cite{Robert:2016a} we can roughly estimate a valley coherence time of around 0.45~ps. This value is of the same order of magnitude as the measured valley coherence time of a highly crystalline CVD-grown WSe$_2$ sample on sapphire substrate \cite{Hao:2015a}.\\
\indent We have shown that the linear polarization axis of the X$^0$ PL rotates as a function of the applied magnetic field. Using the linear polarization basis of the laser also for detection of the PL at $B\neq0$, would lead therefore to a lowering of the observed linear polarization. This is exactly what has been reported recently for ML WSe$_2$ \cite{Aivazian:2015a,Wang:2015d}. 
For the same sample investigated first in Figs.~1d and 2a (where we rotate the detection basis) we show in Fig.~2b the linear polarization degree of the X$^0$ PL for a \textit{fixed} linear basis (parallel to the excitation laser). The linear polarization degree of X$^0$ drops from 15\% at $B=0$ to 7\% at $B\pm$9~T. In our very simple description without considering the magnetic field dependence of the valley coherence generation rate, this linear polarization should be directly deduced by equation (1). In Fig.~2b (fixed detection basis) we generate the solid line with the same parameters as used in Fig.~2a (optimized detection basis), the agreement of the simple model with the experimental results is remarkable. This close fit indicates that changes of the valley coherence generation rate as a function of magnetic field are negligible in our measurement. \\
\textit{Conclusions and Perspectives.---}
Following the demonstration of optically generated valley polarization and valley coherence in the literature, we go a step further by demonstrating the coherent manipulation of valley states. This corresponds to a rotation of the exciton spin around the equator of the Bloch sphere, where the rotation angle is set by the value of the applied magnetic field. This is an important step towards the generation of an arbitrary exciton state in order to reach complete control of exciton states. To access states on the Bloch sphere away from the equator (i.e. $S_z \neq 0$) elliptically polarized light can be used for pseudo-spin initialization. In this case the influence of the long-range electron-hole Coulomb exchange interaction on the exciton pseudo-spin evolution in applied fields needs to be investigated in the future \cite{Maialle:1993a,Yu:2014b,Yu:2014a,Glazov:2014a}.
The clearest signatures of valley coherence in MX$_2$ compounds have been observed in ML WSe$_2$, even using non-resonant excitation \cite{Jones:2013a,Hao:2015a,Wang:2015g}.  Another high quality material with spectrally narrow exciton emission is ML MoSe$_2$, but here very close to resonant excitation is necessary to observe any valley polarization \cite{Kioseoglou:2016a,Wang:2015a}. Only very recently optically generated valley coherence has been observed in PL of acid treated ML MoS$_2$ \cite{Cadiz:2016a}, providing another interesting sample system for valley coherence experiments. \\
\indent \textit{Acknowledgements.---} We thank ERC Grant No. 306719  and ANR MoS2ValleyControl for financial support. X.M. also acknowledges the Institut Universitaire de France. F.C and P. R thank the grant NEXT No ANR-10-LABX-0037 in the framework of the Programme des Investissements d’Avenir”. B. L. acknowledges the support by the National Science Foundation of
China Grant No. 11574357 and the National Basic Research Program of
China Grant No. 2015CB921001. We acknowledge very fruitful discussions with Prof. Juren Shi at early stages of the project.

\end{document}